\documentclass[twocolumn,preprintnumbers,prd,superscriptaddress,nofootinbib,floatfix,showpacs,showkeys]{revtex4-1}
\usepackage{upgreek}

\usepackage{amsmath}
\usepackage{amsfonts}
\usepackage{amssymb}
\usepackage{graphicx}
\usepackage[titletoc]{appendix}
\usepackage[normalem]{ulem}
\usepackage{cancel}
\usepackage{color}
\usepackage[rightcaption]{sidecap}
\usepackage{subfigure}
\usepackage{comment}
\usepackage{mathtools}
\usepackage{dcolumn}
\usepackage{array}
\usepackage{ctable}
\usepackage{multirow}
\usepackage{hhline}
\usepackage{siunitx}
\usepackage{longtable}
\usepackage{tabularx}
\usepackage{booktabs}
\usepackage{adjustbox}

\graphicspath{{Graphics/}{Images/}{Images_old/}{Images_old/recent/}}

\newcommand{\dd}{\mathrm{d}}

\def\bea{\begin{eqnarray}}
\def\eea{\end{eqnarray}}
\newcommand{\onetwo}{\frac{1}{2}}

\newcommand{\PR}{\mathcal{P}_\mathcal{R}}

\newcommand{\MpXi}{\left(M_P^2-\xi\phi^2\right)}

\newcommand{\mpl}{M_{P}}
\newcommand{\Nrem}{N_{\rm rem}}

\definecolor{vividviolet}{rgb}{0.62, 0.0, 1.0}
\definecolor{amaranth}{rgb}{0.9, 0.17, 0.31}
\definecolor{palatinateblue}{rgb}{0.15, 0.23, 0.89}
\definecolor{brightpink}{rgb}{1.0, 0.0, 0.5}
\definecolor{cornflowerblue}{rgb}{0.39, 0.58, 0.93}
\definecolor{deepcarminepink}{rgb}{0.94, 0.19, 0.22}
\definecolor{radicalred}{rgb}{1.0, 0.21, 0.37}

\usepackage{hyperref}
\usepackage{cleveref}
\hypersetup{ linktoc=all,
    colorlinks, linkcolor={palatinateblue},
    citecolor={brightpink}, urlcolor={amaranth}
}
\begin{document}

\title{Non-minimal $\alpha$-attractor magnetogenesis with potential dependent couplings}

\author{Antonino Giacomo Marino}
\email{antonino.marino@unicam.it}
\affiliation{University of Camerino, Via Madonna delle Carceri, Camerino, 62032, Italy}

\author{Tommaso Mengoni}
\email{tommaso.mengoni@unicam.it}
\affiliation{University of Camerino, Via Madonna delle Carceri, Camerino, 62032, Italy}
\affiliation{INAF - Osservatorio Astronomico di Brera, Milano, Italy.}
\affiliation{Istituto Nazionale di Fisica Nucleare (INFN), Sezione di Perugia, Perugia, 06123, Italy.}

\begin{abstract}
We investigate inflationary magnetogenesis driven by a non conformal coupling in the case of large-field $\alpha$-attractor inspired models. We introduce potential-dependent electromagnetic coupling functions $f(\phi)$ that reproduce the scaling $f\propto a^p$ during slow roll. Starting from Starobinsky inflation, we analytically derive explicit couplings for $(E)$-model and $(T)$-model potentials. We then study how these functions affect electromagnetic amplification in the regime of non-minimal coupling between the inflaton field and the Ricci scalar $\phi^2 R$. In particular, we compare different values for the coupling strength $\xi$, accounting for electromagnetic backreaction and Schwinger pair production through the gradient expansion formalism. Our numerical results show that the non-minimal coupling affects the evolution of the electromagnetic sector and the resulting magnetic-field amplitude. We find that, within the analyzed configurations, minimally coupled and $\xi>0$ scenarios are compatible with the adopted CMB constraints, yielding conditional present-day magnetic-field estimates of order $10^{-23}$ G.
\end{abstract}

\keywords{Magnetogenesis, Non-minimal coupling, Inflation, Early Universe, $\alpha$--attractors}

\maketitle

\section{Introduction}

Experimental observations indicate the presence of large-scale magnetic fields in the Universe \cite{collaborationPlanck2015Results2016, tavecchioIntergalacticMagneticField2010,neronovEvidenceStrongExtragalactic2010,andoEvidenceGammaRayHalos2010}, and their origin currently represents an important open problem in modern cosmology \cite{giovannini2004}. 
Different scenarios have been proposed, from astrophysical processes such as, for instance, the Biermann battery \cite{Grasso2001, Attia2021} and dynamo processes, to primordial mechanisms, without guaranteeing a definitive and consistent framework. 

Among primordial scenarios, inflationary magnetogenesis provides a promising theory, in which magnetic fields with large coherence lengths are generated \cite{kandus2011, subramanian2016}. 
In the literature, several approaches have been proposed, see, for example, Refs.\cite{durrer2013,giovannini2004, vachaspati1991,Maiti:2025awl,Maiti:2025cbi,Maiti:2025rkn}.
An interesting scenario consists in introducing a non trivial interaction between the inflaton and the electromagnetic sector, consequently breaking the conformal invariance of the electromagnetic sector \cite{vilchinskii2017}. 
In this respect, a common prescription consists in modifying the electromagnetic Lagrangian through a coupling function $f(\phi)$ that affects the production of primordial magnetic fields and that, after inflation, restores conformal invariance, reproducing standard electromagnetism \cite{Sobol2018}.

The choice of the aforementioned coupling function represents a crucial aspect of the magnetogenesis mechanism. Indeed, different functional forms may significantly modify the electromagnetic amplification, leading to distinct predictions for the generated magnetic fields \cite{martin2008}. In particular, a convenient choice is to assume a power-law dependence on the cosmological scale factor, namely $f\propto a^p$ \cite{ratra1992, Sobol2018}. 

In a previous work \cite{bp3r-s8m2}, inflationary magnetogenesis has been investigated by introducing a non-minimal interaction between the inflaton and the Ricci scalar\footnote{A broader analysis of non-minimally coupled inflation can be found in works such as Refs.~\cite{improvement,hertzberg, density, complete,obscons,futamase,futamase1, LUCCHIN1986163}}. Within this framework, different inflationary potentials and electromagnetic coupling functions have been considered, accounting for electromagnetic backreaction and Schwinger pair production \cite{Sobol2018, Hayashinaka2016a, Stahl2016b, Bavarsad2016a, Kobayashi2014}. In particular, both large- and small-field inflationary scenarios have been studied, showing that the non-minimal coupling parameter can significantly affect the evolution of the electromagnetic sector and the resulting magnetic-field amplitude.

Motivated by these results, in the present work we further investigate the role of the electromagnetic coupling function, focusing on $\alpha$-attractor inspired inflationary models \cite{kallosh, ferrara}. Specifically, starting from the power-law prescription $f\propto a^p$, previously introduced for Starobinsky inflation, we generalize and characterize it to the E- and $T$-model classes of $\alpha$-attractors. Thus, we define specific functions $f(\phi)$ directly related to the corresponding potential $V(\phi)$ and assess whether the obtained different functional forms for $E$-model and $T$-model potentials provide viable scenarios, while preserving the main features of the non-minimal magnetogenesis mechanism.
In this regard, we also employ a recent magnetogenesis approach, such as the gradient expansion formalism \cite{Sobol2020Gradient, Sobol_Gorbar_Teslyk_Vilchinskii_2021, Lysenko2025Gradient}.

The paper is organized as follows. In Sec. \ref{sec 2}, we review the main features of non-minimally coupled inflationary dynamics, while Sec. \ref{sec:electrodynamics} is dedicated to electrodynamics in such inflationary scenarios. In Sec. \ref{sec:couplings}, we introduce the inflationary potentials and derive the corresponding electromagnetic coupling functions. Sec. \ref{sec:numerics} shows the results obtained. Finally, in Sec. \ref{sec concl} we present our conclusions and perspectives.

\section{Non-minimally coupled inflation}\label{sec 2}
\enlargethispage{6pt}

We consider an inflationary scenario in which the inflaton field interacts with both the electromagnetic sector \cite{Sobol2018} and the scalar curvature $R$ \cite{Faraoni2000}, where we employ a non-minimal coupling between the scalar field and the Ricci scalar \cite{Luongo:2024opv, Belfiglio:2024swy, bp3r-s8m2}.

The system is described within the action
\begin{align}\label{tot act}
S=\int&\sqrt{-g}\Biggl[
\frac{1}{2}(M_P^2-\xi\phi^2)R
-\frac{1}{4}f^2(\phi)F^{\mu\nu}F_{\mu\nu}\notag\\
&-\frac{1}{2}g^{\mu\nu}
\partial_\mu\phi\partial_\nu\phi
-V(\phi)+\mathcal{L}_{ch}(A,\theta)
\Biggr]d^4x\,,
\end{align}
where we use metric signature $(-,+,+,+)$ and $M_P$ is the reduced Planck mass, $\xi$ denotes the non-minimal coupling parameter, and $V(\phi)$ represents the inflationary potential. Furthermore, $F_{\mu\nu}=\partial_\mu A_\nu-\partial_\nu A_\mu$
is the electromagnetic tensor, while $f(\phi)$ encodes the interaction between the inflaton and the electromagnetic field. Finally, $\mathcal{L}_{ch}(A,\theta)$ is the Lagrangian density of the charged fields $\theta$ interacting with the vector gauge field $A$.

Non-minimal couplings significantly affect the inflationary evolution.
Specifically, the coupling function $f(\phi)$ breaks the conformal invariance of the electromagnetic sector, allowing for the amplification of electromagnetic fluctuations during inflation.
On the other hand, the interaction term $\xi\phi^2 R$ modifies the gravitational dynamics and, consequently, the evolution of the inflaton field.

Assuming a spatially flat Friedmann-Lema\^itre-Robertson-Walker metric and a homogeneous scalar field, from Eq. \eqref{tot act}, we derive the Klein-Gordon-like equation for the inflaton field
\begin{equation}
\ddot{\phi}+3H\dot{\phi}
+\xi R\phi+V_{,\phi}(\phi)
=-\frac{f_{,\phi}(\phi)f(\phi)}{2}
F^{\mu\nu}F_{\mu\nu}\,,
\label{KG}
\end{equation}
where $H=\dot a/a$ is the Hubble parameter, $R=6(\dot H+2H^2)$ is the Ricci scalar, and commas denote derivatives with respect to the inflaton field.

The modified Einstein equations yield
\begin{equation}\label{efe}
    \MpXi G_{\mu\nu}={T}_{\mu\nu}(\phi,A_\mu)+T^{\theta}_{\mu\nu}\,,
\end{equation}
where the Einstein tensor is defined as $G_{\mu\nu}\equiv R_{\mu\nu}-\onetwo Rg_{\mu\nu}$, $T^{\theta}_{\mu\nu}$ represents the energy-momentum tensor associated with the produced charged particle, while the other contributions are encoded in 
\begin{align}\label{effective tmunu}
    T_{\mu\nu}(\phi,A_\mu)=&f^2(\phi)g^{\alpha\beta}F_{\mu\alpha}F_{\nu\beta}-\frac{1}{4}g_{\mu\nu}f^2(\phi)F^{\alpha\beta}F_{\alpha\beta}\notag\\
    &+\nabla_\mu\phi\nabla_\nu\phi-\onetwo g_{\mu\nu}\nabla^\sigma\phi\nabla_\sigma\phi-g_{\mu\nu}V(\phi)\notag\\&+\xi\left[g_{\mu\nu}\Box\left(\phi^2\right)-\nabla_\mu\nabla_\nu\phi^2\right]\,.    
\end{align}
Consequently, by assuming that $T_{\mu\nu}^\theta$ does not significantly affect the overall trace, we obtain the explicit expressions for the Ricci scalar and the Friedmann equation, respectively
\begin{align}
&R\simeq-\frac{
(1-6\xi)\dot{\phi}^2
-4V(\phi)-6\xi\phi\ddot{\phi}
-18\xi H\phi\dot{\phi}
}{M_P^2-\xi\phi^2}\,,
\label{ricci}\\
&H^2=\frac{1}{3(M_P^2-\xi\phi^2)}
\left(\rho_\phi+\rho_{EM}+\rho_\theta\right)\,,
\label{Fried}
\end{align}
where $\rho_\theta$ is the density of charged particles produced within the Schwinger effect, $\rho_{EM}$ represents the electromagnetic energy contribution. Here we define the inflaton energy density as follows
\begin{equation}
\rho_\phi=
\frac{\dot{\phi}^2}{2}
+V(\phi)+6\xi H\phi\dot{\phi}\,.
\label{rho_phi}
\end{equation}
Notably, the energy density is consistently defined with respect to Eq.\eqref{efe}, but, in general, with different prescriptions, it may present some variations. 

As shown in Eq. \eqref{Fried}, the non-minimal interaction modifies the effective gravitational coupling through the term $M_P^2-\xi\phi^2$, which is required to be positive throughout the inflationary evolution for consistency.
Hence, in this work, in order to conservatively obtain well-defined inflationary scenarios, we focus on the weak non-minimal coupling regime \cite{Luongo:2024opv, Belfiglio:2024swy, bp3r-s8m2}.
To further investigate observational viability for our framework, we evaluate quantities such as the scalar power spectrum $\mathcal P_{\mathcal R}$, the spectral index $n_s$, and the tensor-to-scalar ratio $r$, and confront them with Planck bounds at the pivot scale
$k_*=0.05\,{\rm Mpc}^{-1}$ \cite{planck}.  
In this regard, we consider the weak-coupling expansions retaining terms linear in $\xi$
\begin{align}
     &\mathcal P_{\mathcal R}
    =
    \frac{V^3}
    {12\pi^2\mpl^6(V_{,\phi})^2}
    -\xi
    \frac{
        V^3
        \left(
            8\phi V-3\phi^2V_{,\phi}
        \right)
    }
    {12\pi^2\mpl^8(V_{,\phi})^3}
    +\mathcal O(\xi^2),
\label{eq:PRlinear}\\
 & n_s
    =
    1+
    \mpl^2
    \frac{
        -3(V_{,\phi})^2+2VV_{,\phi\phi}
    }{V^2}+\xi
    \frac{
        8V^2
        -10\phi VV_{,\phi}
    }{V^2}\notag\\
    &\qquad\,+\xi
    \frac{
        3\phi^2(V_{,\phi})^2
        -2\phi^2VV_{,\phi\phi}
    }{V^2}
    +\mathcal O(\xi^2),
\label{eq:nslinear}\\
    &r
    =
    8\mpl^2
    \frac{(V_{,\phi})^2}{V^2}
    +8\xi
    \frac{
        V_{,\phi}
        \left(
            8\phi V-\phi^2V_{,\phi}
        \right)
    }{V^2}
    +\mathcal O(\xi^2).
\label{eq:rEinstein}
\end{align}
We remark that this expansion has been computed in both Jordan and Einstein frames, resulting in an equivalence for these observational parameters at first-order in approximation.
The detailed derivation is presented in Refs.~\cite{postma, bp3r-s8m2}.  
In particular, the limit $\xi\to0$ in Eqs.~\eqref{eq:PRlinear}--\eqref{eq:rEinstein} reduces to the standard canonical expressions.

\section{Electrodynamics during inflation}
\enlargethispage{6pt}
\label{sec:electrodynamics}

In the kinetically coupled model, the Maxwell equations in the presence of a charged current read
\begin{equation}
    \frac{1}{\sqrt{-g}}
    \partial_\mu\left[\sqrt{-g}\,f^2(\phi)F^{\mu\nu}\right]
    =-j^\nu.
    \label{eq:Maxwell}
\end{equation}
In the present work, we adopt the Coulomb gauge and denote by $\mathbf E$ and $\mathbf B$ the physical fields measured by a comoving observer\footnote{Explicitly, we define 
\begin{equation}
    E_i = -\frac{F_{0i}}{a} \,,\qquad B _i =\frac{ F_{ij}\epsilon_{ijk}}{2a^2}\,.
\end{equation}
}.  

In the limit of negligible current source, the mode equation is simply written as the evolution equation for each helicity mode, $\mathcal A_\lambda=f(\phi)A_\lambda$ 
\begin{equation}
    \mathcal A''+
    \left(k^2-\frac{f''}{f}\right)\mathcal A=0,
    \label{eq:modes}
\end{equation}
where the Bunch--Davies vacuum condition \cite{Bunch:1978yq} is applied deep inside the horizon
\begin{equation}
    \mathcal A(\eta,k)=\frac{1}{\sqrt{2k}}e^{-ik\eta},
    \qquad k\eta\rightarrow  -\infty.
    \label{eq:BD}
\end{equation}

Let us introduce the family of auxiliary functions
\begin{equation}
    f_\ell(t)
    \equiv
    \frac{1}{fH^\ell}\frac{\dd^\ell f}{\dd t^\ell},
    \qquad \ell\in\mathbb N.
    \label{eq:fl}
\end{equation}
Since $f''/f=a^2H^2(f_1+f_2)$, the following condition defines the pump-dominated regime; tachyonic instability additionally requires $f_1+f_2>0$ 
\begin{equation}
    k^2\ll a^2H^2\left|f_1+f_2\right|.
    \label{eq:functionalSH}
\end{equation}
When such a condition holds, we state that the mode with comoving wavenumber $k$ is functionally super-Horizon.  We point out that this condition is distinct from the Hubble criterion $k\ll aH$, and it naturally arises from the electromagnetic dynamics.  Now, at a given moment in time, the largest mode that has crossed the functional horizon is defined by the monotonic upper envelope
\begin{align}
    &k_h(t)
    =
    {\rm Env}^{\uparrow}
    \left\{aH\sqrt{|f_1+f_2|}\right\},\notag\\
    &{\rm Env}^{\uparrow}\{F\}(t)
    =\max_{\tau\in[t_i,t]}F(\tau).
    \label{eq:kh}
\end{align}
This construction identifies the largest comoving wavenumber that has crossed the functional horizon up to a given time, even when the corresponding horizon scale evolves nonmonotonically \cite{Sobol2020Gradient}.
Introducing $k_h(t)$ is particularly useful for keeping track of the energy transfer due to the modes crossing the horizon during the cosmological expansion. Indeed, for any spectral moment $\mathcal X^{(n)}$, the flux of newly relevant modes crossing the functional horizon (and hence contributing to the coarse grained electromagnetic sector) is
\begin{equation}
    \left[\dot{\mathcal X}^{(n)}\right]_b
    =
    \left.
    \frac{\dd\mathcal X^{(n)}}{\dd\ln k}
    \right|_{k=k_h(t)}
    \frac{\dd\ln k_h}{\dd t}.
    \label{eq:boundaryGeneral}
\end{equation}
Explicitly, the spectra entering Eq.\eqref{eq:boundaryGeneral} are evaluated with the Hankel approximation to Eq.~\eqref{eq:modes} near crossing.  As a particular example, let us compute the electric-energy injection, for which we write 
\begin{equation}
        \left[\dot{\rho}_E\right]_b
    =
    \left.
    \frac{\dd{\rho}_E}{\dd\ln k}
    \right|_{k=k_h(t)}
    \frac{\dd\ln k_h}{\dd t}.
\end{equation}
Here, we define the normalized potential time derivative
\begin{equation}
    \mathcal D(\eta,k)
    \equiv
    \frac{f}{k}\frac{\partial}{\partial\eta}
    \left(\frac{\mathcal A}{f}\right),
    \label{eq:Ddef}
\end{equation}
in terms of which %
the electric energy spectrum can be expressed, resulting in the following horizon source term
\begin{equation}
    \left(\dot\rho_E\right)_b
    =
    \frac{1}{2\pi^2}\frac{k_h^{5}}{a^4}
    \left|\mathcal D(\eta,k_h)\right|^2
    \frac{\dd\ln k_h}{\dd t},
    \label{eq:boundaryD}
\end{equation}
Specifically, $\mathcal D$ can be analytically computed near horizon crossing
\begin{equation}
\begin{split}
    \mathcal D(\eta,k)
    &=
    \frac{
    e^{i\left(\nu_h\frac{\pi}{2}+\frac{\pi}{4}\right)}
    }{2\sqrt{2k}}
    \sqrt{\frac{\pi}{2}(-k\eta)}
   \Bigg[
    \left(1-\frac{\frac12+f_1(t)}{\nu_h}\right)\times
    \\
    &
    \times H_{\nu_h+1}^{(1)}(-k\eta)-\left(1+\frac{\frac12+f_1(t)}{\nu_h}\right)
    H_{\nu_h-1}^{(1)}(-k\eta)
    \Bigg].
\end{split}
\label{eq:Danalytic}
\end{equation}
where the adopted real-index prescription is 
\begin{equation}
        \nu_h
    =\sqrt{\left|\frac14+f_1(t_h)+f_2(t_h)\right|},
\end{equation}
and $H^{(1)}$ is the Hankel function of the first kind. The absolute-value prescription coincides with the local Hankel index of \cite{Sobol2020Gradient} when $1/4+f_1(t_h)+f_2(t_h)\geq0$; for a negative radicand it modifies the local Hankel approximation.

\subsection{Including electromagnetism in the numerical background}\label{sec 2.2.1}

The backreaction problem requires accounting for the electric and magnetic energy contribution to the Friedmann equation. %
However, solving the coupled background equations with the evolution of a large number of
electromagnetic Fourier modes is computationally
demanding.
One possible approach to tackle this issue involves the assumption that the magnetic energy density is approximately negligible with respect to the electric counterpart during the inflationary evolution. In this scenario, the Maxwell equation is translated into a continuity equation for the electric density 
\begin{equation}
    \dot\rho_E+4H\rho_E
    +2\frac{\dot f}{f}\rho_E
    =\left(\dot\rho_E\right)_b-Q_{\rm Schw},
    \label{eq:rhoEfull}
\end{equation}
where the Schwinger particle production is entirely due to the electric decay and the associated source term is written as
\begin{equation}
    Q_{\rm Schw}
    \equiv\mathbf E\cdot\mathbf J_{\rm Schw}.
    \label{eq:SchwingerSourceCA}
\end{equation}
Here $\mathbf J_{\rm Schw}$ is the physical current measured by a comoving observer. This same energy is assumed to be entirely transferred to an ultrarelativistic fluid of charged particles, whose density satisfies
\begin{equation}
    \dot\rho_\theta+4H\rho_\theta=Q_{\rm Schw}.
    \label{eq:rhoCharged}
\end{equation}
A second possible approach for accounting for the electromagnetic influence at background level is represented by the Gradient Expansion Formalism (GEF). This approach replaces the spatial dependence of the long-wavelength electromagnetic field by an infinite hierarchy of homogeneous bilinears \cite{Sobol2020Gradient, Lysenko2025Gradient}. In this framework, the Maxwell equation is recast as an infinite series of ODEs in the variables
\begin{align}
    \mathcal E^{(n)}
    &\equiv
    \frac{f^2}{a^n}
    \left\langle
    \mathbf E\cdot(\boldsymbol\nabla\!\times)^n\mathbf E
    \right\rangle,
    \\
    \mathcal G^{(n)}
    &\equiv
    -\frac{f^2}{2a^n}
    \left\langle
    \mathbf E\cdot(\boldsymbol\nabla\!\times)^n\mathbf B
    +(\boldsymbol\nabla\!\times)^n\mathbf B\cdot\mathbf E
    \right\rangle,
    \\
    \mathcal B^{(n)}
    &\equiv
    \frac{f^2}{a^n}
    \left\langle
    \mathbf B\cdot(\boldsymbol\nabla\!\times)^n\mathbf B
    \right\rangle.
    \label{eq:GEFmoments}
\end{align}
Here, we denote the corresponding Schwinger-scale-filtered moments by $\overline{\mathcal E}^{(n)}$, $\overline{\mathcal B}^{(n)}$, and $\overline{\mathcal G}^{(n)}$ (more details in Sec.~\ref{sec:schwinger}) and obtain the GEF hierarchy as

\begin{align}
    \dot{\mathcal E}^{(n)}
    &+(n+4)H\mathcal E^{(n)}
    +2\frac{\dot f}{f}\mathcal E^{(n)}
    +2\mathcal G^{(n+1)}
    +2\frac{\sigma_E}{f^2}\overline{\mathcal E}^{(n)}\notag\\&
    -2\frac{\sigma_B}{f^2}\overline{\mathcal G}^{(n)}
    =\left[\dot{\mathcal E}^{(n)}\right]_b,
    \label{eq:GEFE}\\
    \dot{\mathcal B}^{(n)}
    &+(n+4)H\mathcal B^{(n)}
    -2\frac{\dot f}{f}\mathcal B^{(n)}
    -2\mathcal G^{(n+1)}
    =\left[\dot{\mathcal B}^{(n)}\right]_b,
    \label{eq:GEFB}\\
    \dot{\mathcal G}^{(n)}
    &+(n+4)H\mathcal G^{(n)}
    +\mathcal B^{(n+1)}-\mathcal E^{(n+1)}
    +\frac{\sigma_E}{f^2}\overline{\mathcal G}^{(n)}
    \notag\\&-\frac{\sigma_B}{f^2}\overline{\mathcal B}^{(n)}
    =\left[\dot{\mathcal G}^{(n)}\right]_b.
    \label{eq:GEFG}
\end{align}
Eqs. \eqref{eq:GEFE}-- \eqref{eq:GEFG} represent a recursive system where the variables associated with the index $n$ can only be solved once the index $n+1$ has been integrated. In other words, solving the electrodynamics with exact precision requires the integration of a system with an infinite number of ODEs. Yet, previous works have shown that a good approximation can be obtained by truncating this recursive series at a sufficiently high order, and in this work we stop at $n_{\max}=15$. The truncation of the system can be achieved by imposing closure with the ultraviolet scale selected by the functional horizon,
\begin{equation}
    \mathcal X^{(n+2)}
    \simeq
    \left(\frac{k_h}{a}\right)^2\mathcal X^{(n)},
    \qquad
    \mathcal X\in\{\mathcal E,\mathcal B,\mathcal G\}.
    \label{eq:GEFclosure}
\end{equation}
Truncation at an odd maximum mixed order can be checked against the next odd order and Fourier modes on the same background. For the parity-symmetric kinetic-coupling case, we use the mode equation of \cite{Lysenko2025Gradient}, with vanishing axial coupling and magnetic conductivity:
\begin{align}
\ddot{\mathcal A}_{\lambda}
&+\left(H+\frac{\sigma_E\Theta}{f^2}\right)\dot{\mathcal A}_{\lambda}\notag\\&
+\left[\frac{k^2}{a^2}
-\frac{\ddot f+(H+\sigma_E\Theta/f^2)\dot f}{f}\right]\mathcal A_{\lambda}=0.
\label{eq:modeSchwinger}
\end{align}
Here $\Theta(t,k)$ is the filter in Eq.~\eqref{eq:schwinger-theta}, including the production-scale activation used in our implementation; the published mode filter is $\theta[k_h(t)-k]$. The conductivity and background are supplied by the GEF solution, so this is an internal consistency test. 
The GEF-Schwinger prescription adopted in this work provides a numerical framework for consistently incorporating the backreaction of electromagnetic field production and Schwinger pair creation into the inflationary dynamics\footnote{
Within this framework, the Schwinger source term appearing in Eq.~\eqref{eq:rhoCharged} takes the form
\begin{equation}
    Q_{\rm Schw}
    =\frac{1}{f^2}
    \left(
        \sigma_E\overline{\mathcal E}^{(0)}
        -\sigma_B\overline{\mathcal G}^{(0)}
    \right),
    \label{eq:QSchwGEF}
\end{equation}
while the electromagnetic energy densities entering the Einstein equations are expressed as
\begin{equation}
    \rho_E=\frac12\mathcal E^{(0)},
    \qquad
    \rho_B=\frac12\mathcal B^{(0)}.
    \label{eq:GEFenergies}
\end{equation}
}. Nevertheless, we emphasize that the applicability of this prescription is restricted to the inflationary epoch. In particular, its direct extension to the reheating phase would not provide a sufficiently accurate description of the electromagnetic dynamics, as the emergence of a plasma of charged particles and the associated conductive currents necessitate a more comprehensive treatment of the electromagnetic sector, including the evolution and possible perturbations of the electrical conductivity.

\subsection{Schwinger source}
\label{sec:schwinger}

We also include particle--antiparticle production through the Schwinger effect. Pair production and the induced current in de Sitter spacetime have been studied for scalar and fermionic matter by \cite{Kobayashi2014, Hayashinaka2016a,Hayashinaka2016b, Stahl2016a, Stahl2016b}, while backreaction and nonequilibrium formulations of the current have a longer history \cite{Shaisultanov_1995}. In the context of inflationary gauge-field production, the interplay among the Schwinger effect, chiral or axial interactions, and gradient-expansion dynamics has been developed by \cite{Domcke_Ema_Mukaida_2020,Domcke_Ema_Mukaida_2021,Gorbar_Schmitz_Sobol_Vilchinskii_2021a,Gorbar_Schmitz_Sobol_Vilchinskii_2022}. Here we adopt the mixed electric--magnetic, scale-dependent description of \cite{vonEckardstein_Schmitz_Sobol_2025}. Reconstructing the field invariants from the zeroth-order bilinears, we obtain
\begin{equation}\label{eq:gef-physical-invariants}
E^2=\frac{\mathcal E^{(0)}}{f^2},
\qquad
B^2=\frac{\mathcal B^{(0)}}{f^2},
\qquad
P\equiv{\bf E\!\cdot\! B}=-\frac{\mathcal G^{(0)}}{f^2}.
\end{equation}
We further define 
\begin{align}
\Delta&=\sqrt{(E^2-B^2)^2+4P^2},\nonumber\\
E_\parallel&=\sqrt{\frac{E^2-B^2+\Delta}{2}},\label{eq:parallel-invariants}\\
B_\parallel&=\sqrt{\frac{B^2-E^2+\Delta}{2}},\nonumber
\end{align}
where $E_\parallel$ and $B_\parallel$ are the non-negative magnitudes of the fields in the local Lorentz frame in which the electric and magnetic fields are collinear. 

For collinear electric and magnetic fields, the Schwinger current is aligned with both fields, so its decomposition into electric and magnetic contributions is not unique. This ambiguity matters when the fields are promoted to operators while the conductivities are treated as classical coefficients. Following \cite{vonEckardstein_Schmitz_Sobol_2025}, we adopt the mixed prescription obtained by considering non-collinear fields: the current is evaluated in a local Lorentz frame where the fields are collinear and then transformed back to the comoving frame. This procedure fixes the physical conductivities $\sigma_E$ and $\sigma_B$ appearing in Eqs.~\eqref{eq:GEFE}--\eqref{eq:GEFG}. They are evaluated from the mean-field invariants as classical coefficients, yielding
\begin{align}
\sigma_E={}&\frac{C g'^3}{6\pi^2H}
\frac{|P|}{\sqrt{\Delta}}
\sqrt{\frac{E^2-B^2+\Delta}{E^2+B^2+\Delta}}
\coth\!\left(\frac{\pi B_\parallel}{E_\parallel}\right),\label{eq:mixed-conductivities0}\\
\sigma_B={}&\frac{C g'^3}{6\pi^2H}
\frac{P}{\sqrt{\Delta}}
\sqrt{\frac{B^2-E^2+\Delta}{E^2+B^2+\Delta}}
\coth\!\left(\frac{\pi B_\parallel}{E_\parallel}\right).
\label{eq:mixed-conductivities}
\end{align}
Here, $C$ is the species factor accounting for effectively massless Standard Model hypercharge carriers and is evaluated as $C=41/12$. In our description we also consider a scale invariant weak-hypercharge gauge coupling $g'\simeq0.35$, hence we neglect its renormalization-group running. A detailed analysis on the effects on this model of $g'$ being a function of the energy scale goes beyond the scope of the present work and is left for future works.

In the standard picture, the conductive medium is assumed to be scale dependent. In this regard, we employ the simplified prescription which expects the Schwinger effect to be significant only in the range of momenta that runs up to the monotonic functional-horizon envelope $k_h$. Thus, we assume the conductivities to be scale dependent as follows
\begin{equation}
    \sigma_{E,B}(t,k)=\sigma_{E,B}(t)\Theta(t,k)\,,
\end{equation}
where
\begin{equation}\label{eq:schwinger-theta}
\Theta(t,k)=\theta\!\bigl(k_h(t)-k\bigr)\,
\theta\!\bigl(k_S(t)-a(t)H(t)\bigr).
\end{equation}
The second step function activates the local production approximation only when $k_S/a>H$, so its characteristic length is shorter than the Hubble radius. Given the Standard model species factors defined before, the Schwinger scale is evaluated from the physical electric field in the collinear frame as follows
\begin{equation}\label{eq:schwinger-scale}
k_S(t)=\frac{a(t)C^{1/3}|g'|}{2^{1/4}}
\left[E^2-B^2+\Delta\right]^{1/4}
=a(t)C^{1/3}|g'|\sqrt{E_\parallel}.
\end{equation}
Eq. \eqref{eq:schwinger-theta} is defined in such a way that once $k_S>aH$, all modes represented by the GEF sector are subject to the Schwinger damping terms, whereas for $k_S\leq aH$ the conductive terms vanish. As a consequence, no scale dependent variables are evolved in this approximation, and the GEF system retains a much simpler form. 

With this filter, $\overline{\mathcal X}^{(n)}=\theta(k_S-aH)\mathcal X^{(n)}$ for $\mathcal X\in\{\mathcal E,\mathcal B,\mathcal G\}$ in Eqs.~\eqref{eq:GEFE}--\eqref{eq:GEFG} and \eqref{eq:QSchwGEF}. The parity-symmetric numerical implementation has $\mathcal G^{(0)}=0$ and $\sigma_B=0$, and evolves even electric/magnetic moments and odd mixed moments.\footnote{In the electric-dominated regime, the $P\to0$ limit of the conductivity is evaluated continuously, using $B_\parallel\coth(\pi B_\parallel/E_\parallel)\to E_\parallel/\pi$ when $B_\parallel\to0$.}  

\subsection{Magnetic spectrum and present-day field}

It is possible to obtain an expression for the maximal value that the produced magnetic field can retain at present day epoch. This extremal evaluation can be estimated by assuming the magnetic density to dilute like the fourth power of the scale factor for the duration of the $N\sim 65 $ e-folds separating the present day from the end of the inflationary process. In this treatment, a cutoff on the power spectrum is assumed, to account for the complete dilution of those mode whose characteristic time scale results smaller than the current age of the Universe \cite{Grasso2001}.

Explicitly, the magnetic power spectrum is 
\begin{equation}
    \frac{\dd\rho_B}{\dd k}
    =\frac{k^4}{2\pi^2a^4}|\mathcal A(\eta,k)|^2,
\label{eq:PB}
\end{equation}
and assuming the standard post-inflationary magnetic dilution, the following conditional maximum-survival estimate is obtained for the present-day magnetic-field amplitude 
\begin{equation}
    B_0
    =
    \left(\frac{a_e}{a_0}\right)^2
    \left[
        2\int^{k_{\rm diff}}
        \frac{\dd k}{k}
        \left.
        \frac{\dd\rho_B}{\dd\ln k}
        \right|_{t_e}
    \right]^{1/2},
    \label{eq:B0}
\end{equation}
where $k_{\rm diff}$ is the comoving diffusion cutoff. We use $k_{\rm diff}/a_0=10^{-46}M_P\simeq2.44\times10^{-28}\,{\rm GeV}$, which selects modes surviving diffusion until the present day.

\section{Modeling the electromagnetic coupling functions}%
\label{sec:couplings}

In this section we explicitly address the choice for the inflaton's potential and the consequent construction of the kinetic coupling function. As anticipated, we target the class of $\alpha$-attractor potentials, corresponding to the Starobinsky model, the $E$-model and the T-model. 
The reconstruction of the function $f(\phi) \sim a^p$ will be achieved by means of an analytical integration of the inflaton's evolution in the slow-roll regime for each of the potentials. The coupling is normalized by $f(0)=1$, at the potential minimum, and decreases along the positive-field inflationary branch in order to avoid the strong coupling problem \cite{demozzi2009}.

Here, we assume to be working in the limit $\xi\rightarrow 0$, as required by the present prescription to keep the coupling function $f$ independent of the strength parameter $\xi$.
In this framework, it is possible to express the magnitude of the scale factor as a function of the chosen potential $    \frac{\dd\ln a}{\dd\phi}
    \simeq
    -\frac{V}{\mpl^2V_{,\phi}}$, where within the slow-roll approximation, Eqs. \eqref{KG} and \eqref{Fried} read
\begin{equation}
    3H\dot\phi\simeq-V_{,\phi},
    \qquad
    3\mpl^2H^2\simeq V.
\end{equation} In this way, it is possible to use the nominal value of the inflaton field as a variable indicating the inflationary evolution. Explicitly, we write the remaining number of e-folds between a given moment in time and the end of inflation as
\begin{equation}
    \Nrem(\phi)
    =
    \frac{1}{\mpl^2}
    \int_{\phi_{\rm end}}^\phi
    \frac{V}{V_{,\phi}}\,\dd\phi.
    \label{eq:NremGeneral}
\end{equation}
Finally, by virtue of the definition of e-fold number, we write the candidate form for the coupling function as
\begin{equation}
 f(\phi)=\exp\!\left[-\frac{p}{M_P^2}\int_0^\phi
 \frac{V(\varphi)}{V_{,\varphi}(\varphi)}\,\dd\varphi\right].
 \label{eq:fGeneral}
\end{equation}
where the conformal restoration condition $f(0)=1$ is enforced by subtracting $\Nrem(0)$, rather than normalizing at the end of inflation.
In the present work we conventionally use $p=-5/2$, so $f>1$ during inflation and approaches unity from
above.
We consider as starting point the Starobinsky model \cite{starob}, and then the $E$- and $T-$model classes for $\alpha$-attractor potentials \cite{kallosh, ferrara}
\begin{align}
   V_{\rm St}(\phi)
    &=
    \Lambda_{\rm St}^4
    \left[
        1-\exp\!\left(
        -\sqrt{\frac23}\frac{\phi}{\mpl}
        \right)
    \right]^2,
    \label{eq:VSt}\\
    V_E^{(n)}(\phi)
    &=
    \Lambda_E^4
    \left[
        1-\exp\!\left(
        -\sqrt{\frac{2}{3\alpha_E n}}
        \frac{\phi}{\mpl}
        \right)
    \right]^{2n},
    \label{eq:VE}\\
     V_T^{(m)}(\phi)
    &=
    \Lambda_T^4
    \tanh^{2m}\!\left(
        \sqrt{\frac{1}{6\alpha_Tm}}
        \frac{\phi}{\mpl}
    \right),
    \label{eq:VT}
\end{align}
where $\Lambda_i$ and $\alpha_i$ are free parameters of the potentials.
The analytical derivation of the corresponding coupling functions is kept general in $n$ and $m$. The
numerical section later fixes $n=2$ and $m=1$ as benchmarks for the analysis.

A direct integration of Eq. \eqref{eq:fGeneral} for each of these cases produces the corresponding coupling functions %
\begin{align}
    f_{\rm St}(\phi)
    &=
    \exp\!\left\{
    -\frac{3p}{4}
    \left[
        \exp\!\left(\sqrt{\frac23}\frac{\phi}{M_P}\right)
        -
        1
        -
        \sqrt{\frac23}\frac{\phi}{M_P}
    \right]
    \right\}
    ,
    \label{eq:fSt}\\
    f_E(\phi)
    &=
    \exp\!\Biggl\{
    -\frac{3p\alpha_E}{4}
    \Biggl[
        \exp\!\left(\sqrt{\frac{2}{3\alpha_E n}}\frac{\phi}{M_P}\right)
        -
        1
        \notag\\&\qquad\quad\,\,-
        \sqrt{\frac{2}{3\alpha_E n}}\frac{\phi}{M_P}
    \Biggr]
    \Biggr\}
    ,
    \label{eq:fEpaper}\\
    f_T(\phi)
    &=
    \exp\!\left\{
    -\frac{3p\alpha_T}{4}
    \left[
        \cosh(\sqrt{\frac{2}{3\alpha_T m}}\frac{\phi}{M_P})-1
    \right]
    \right\}
    ,
    \label{eq:fTpaper}
\end{align}
where the coupling function for the Starobinsky model in Eq. \eqref{eq:fSt} is already used in works such as, for instance, Refs. \cite{Sobol2018, bp3r-s8m2}.

\section{Numerical setup}
\label{sec:numerics}

The numerical methods developed in this instance aim to assess the capability of the introduced models in matching the primordial cosmological constraints. In particular, we develop a numerical optimization engine designed to identify, for each model, the specific configuration of free parameters whose predicted observables $\PR$,  $ r$ and $n_s$ fall inside the boundaries outlined by the results of the Planck mission \cite{planck}.

This numerical workflow separates the optimization of the parameter selection from the final physical evaluation of the model. Specifically, we organized the treatment in three different steps:
\begin{enumerate}
    \item  \textbf{The parameter calibration}. In order to calibrate the parameter configuration, we integrate Eqs. \eqref{Fried}, \eqref{KG}, \eqref{eq:rhoEfull} and \eqref{eq:rhoCharged} with an adaptive Runge--Kutta method and event location at the first rising solution of $\epsilon_V\equiv M_P^2(V_{,\phi}/V)^2/2=1$. This is the potential slow-roll endpoint used by the saved runs, rather than the exact acceleration endpoint $\epsilon_H=-\dot H/H^2=1$. Essentially, we do not make use of the full GEF for this preliminary stage, as the continuity equation approach to the electric backreaction is less computationally demanding but still able to provide an approximated result consistent with the final GEF validation.
We impose the following initial conditions on every model
\begin{align}\label{eq:init_cond}
    &a(t_i)=1,\qquad
    \dot\phi(t_i)=-\frac{M_P\,V_{,\phi}(\phi_i)}{\sqrt{3V(\phi_i)}},\notag\\
    &\rho_E(t_i)=\rho_\theta(t_i)=10^{-30}\mpl^4.
\end{align}

The free parameters to be optimized are $\phi_i,\Lambda_{\rm i},N_*$, namely the initial condition of the inflaton field, the scale of the inflaton's potential and the number $N_*$, which marks the e-fold count between the pivot scale crossing and the end of inflation. Only for the $E$-model and T-model, another further free parameter is added in the optimization picture, respectively $\alpha_E$ and $\alpha_T$, which mark the transition sharpness of the two potentials. The optimization screening marks with failures those trials that violate the model domain $r<0.06$, fail the Friedmann constraint, do not end inflation, or lie outside $60\lesssim N_{\rm end}\lesssim75$. In this picture, the cost function is defined as a weighted quadratic deviation from the observables central value  $\PR =2.101\times10^{-9}\,,\,\, n_s=0.9651$. Finally, we fix $\xi\in\{\pm5\cdot10^{-4},0\}$, $p=-5/2$, $n=2$ and $m=1$.

\item \textbf{The GEF integration of the selected points.} For every selected potential--$\xi$ pair, the optimized
$\{\phi_i,\Lambda_i,\alpha_i,N_*\}$ subset is passed unchanged to the GEF solver together with the corresponding coupling in Eqs.~\eqref{eq:fSt}, \eqref{eq:fEpaper}, or \eqref{eq:fTpaper}. The scalar, scale-factor, and charged-fluid initial conditions of Eq.~\eqref{eq:init_cond} are retained, while the electromagnetic hierarchy starts from zero coarse-grained moments, with subsequent modes injected through Eq.~\eqref{eq:boundaryGeneral}. In this stage the Schwinger sector is recomputed from the scale-filtered GEF bilinears using the disambiguated conductivities in Eqs.~\eqref{eq:mixed-conductivities0} and \eqref{eq:mixed-conductivities} at maximum mixed order $n_{\max}=15$, with an $n_{\max}=17$ convergence comparison. 
\item \textbf{The final viability checks.} We close every numerical routine monitoring that the behavior of the following variables behaves according to the analytical prediction, paying a particular attention to the consistency with the negligible $\rho_B$ assumption (necessary for obtaining Eq. \eqref{eq:rhoEfull}) 
\begin{align}
    &\frac{\rho_E}{\rho_\phi}<1,\qquad
    \frac{\rho_B}{\rho_E}\ll1,\qquad
    \frac{\rho_\theta}{\rho_\phi}\ll1,\notag\\&
    \frac{\dd\ln f}{\dd\ln a}\simeq p,\qquad
    \frac{k_h}{aH}\sim \text{const}.
\end{align}
\end{enumerate}
Finally, the values of $\mathcal P_{\mathcal R}$, $n_s$, and $r$ are recomputed at the GEF pivot point and subjected to the same rectangular Planck test used in Stage~1. The conditional present-day field $B_0$ is computed from the diffusion-filtered spectrum of independently evolved Fourier modes on the GEF background, without feeding those modes back into the background.
After the adopted inflation endpoint, we enforce $\rho_B(t)=\rho_B(t_e)[a_e/a(t)]^4$ to depict a conditional upper bound on the surviving magnetic energy density in the absence of additional amplification. The plotted electric density is continued with the same $a^{-4}$ prescription. These dashed continuations are imposed dilution laws, not reheating solutions. Indeed, the inflationary GEF--Schwinger conductivity prescription is not reliable during reheating, when the charged-particle distributions, plasma interactions, and conductivities change. 

\begin{figure*}[ht]
    \centering
    \includegraphics[width=1\linewidth]{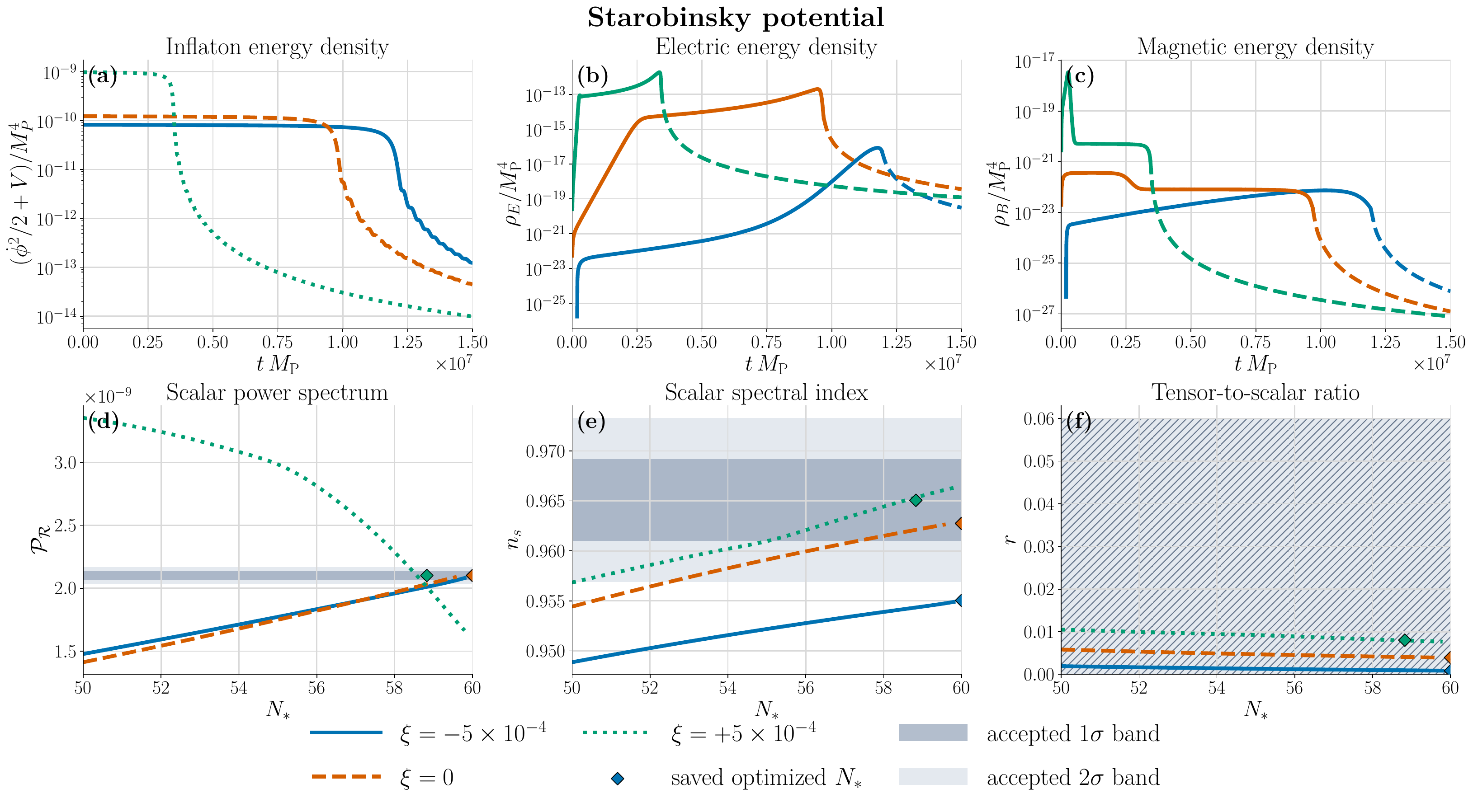}
    \caption{Starobinsky potential with the potential-adapted coupling. Panels (a)--(c) show the inflaton, electric, and magnetic energy densities, respectively, as functions of time in reduced-Planck units. Panels (d)--(f) show the scalar curvature power spectrum $\mathcal P_{\mathcal R}$, scalar spectral index $n_s$, and tensor-to-scalar ratio $r$, respectively, as functions of the pivot-crossing e-fold number $N_*$. Panel (a) shows the canonical scalar contribution $\dot\phi^2/2+V$, excluding the $6\xi H\phi\dot\phi$ term in Eq.~\eqref{rho_phi}. In panels (b)--(c), each color identifies one $\xi$; solid lines precede the saved $\epsilon_V=1$ endpoint and dashed lines show the imposed post-inflationary dilution. Diamond markers in panels (d)--(f) denote the continuity-approximation calibration values.}
    \label{fig:STAR}
\end{figure*}
\subsection{Numerical results}
\label{sec:results}
\begin{table}[ht]
\centering
\caption{Observable bounds used for the shaded accepted regions in Figs.~\ref{fig:STAR}--\ref{fig:TMOD}. Displayed bounds are rounded; the plotted regions and acceptance tests use the unrounded values. The tensor-to-scalar ratio has the same one-sided upper bound in both regions.}
\label{tab:observable_bounds}
\begingroup
\begin{tabular}{@{}ccc@{}}
\toprule
\hline
Observable & Region & Accepted range or upper bound \\\hline
\midrule
$\mathcal P_{\mathcal R}$ & $1\sigma$ & $[2.07,\,2.14]\times10^{-9}$ \\
$\mathcal P_{\mathcal R}$ & $2\sigma$ & $[2.03,\,2.17]\times10^{-9}$ \\
\addlinespace[3pt]
$n_s$ & $1\sigma$ & $[0.961,\,0.969]$ \\
$n_s$ & $2\sigma$ & $[0.957,\,0.973]$ \\
\addlinespace[3pt]
$r$ & $1\sigma$ & $r\leq0.06$ \\
$r$ & $2\sigma$ & $r\leq0.06$ \\
\bottomrule
\hline
\end{tabular}
\endgroup
\end{table}
\begin{table}[ht]
\centering
\caption{Fitted potential parameters for the analyzed configurations, in reduced-Planck units. Here, $\alpha_i$ denotes $\alpha_E$ or $\alpha_T$ for the corresponding model; it does not apply to the Starobinsky potential. }
\label{tab:fitted_parameters}
\begingroup

\begin{tabular}{@{}crccc@{}}
\toprule
\hline
Model & $\xi\qquad$ & $\quad\phi_i / M_P$ \quad& $\quad\Lambda_i/M_P\quad$ & $\quad\alpha_i\quad$ \\
\midrule
\hline
Starobinsky & $\quad-5\times10^{-4}$ & $5.17$ & $0.00303$ & N.A. \\
Starobinsky & $0$ & $5.28$ & $0.00335$ & N.A. \\
Starobinsky & $5\times10^{-4}$ & $5.27$ & $0.00562$ & N.A. \\
\addlinespace[3pt]
$E$-model & $-5\times10^{-4}$ & $8.16$ & $0.00397$ & $1.46$ \\
$E$-model & $0$ & $9.51$ & $0.00499$ & $2.50$ \\
$E$-model & $5\times10^{-4}$ & $9.25$ & $0.00651$ & $2.19$ \\
\addlinespace[3pt]
$T$-model & $-5\times10^{-4}$ & $9.36$ & $0.00464$ & $5.00$ \\
$T$-model & $0$ & $8.90$ & $0.00490$ & $4.17$ \\
$T$-model & $5\times10^{-4}$ & $6.96$ & $0.00615$ & $1.39$ \\
\bottomrule
\hline
\end{tabular}
\par\medskip
\endgroup
\end{table}

\begin{figure*}[t]
    \centering
    \includegraphics[width=1\linewidth]{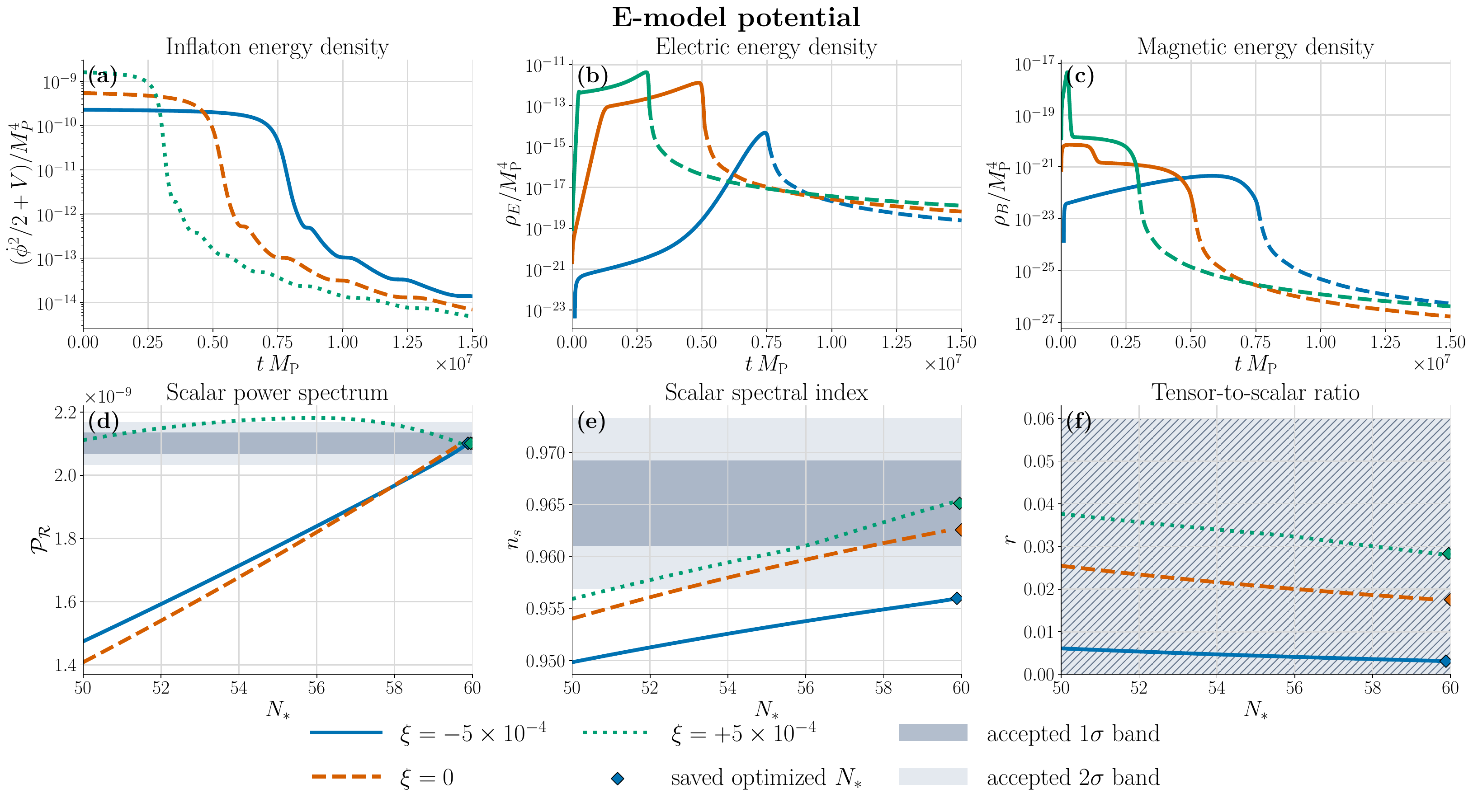}
    \caption{$E$-model potential with the potential-adapted coupling. Panels (a)--(c) show the inflaton, electric, and magnetic energy densities, respectively, as functions of time in reduced-Planck units. Panels (d)--(f) show the scalar curvature power spectrum $\mathcal P_{\mathcal R}$, scalar spectral index $n_s$, and tensor-to-scalar ratio $r$, respectively, as functions of the pivot-crossing e-fold number $N_*$. Panel (a) shows the canonical scalar contribution $\dot\phi^2/2+V$, excluding the $6\xi H\phi\dot\phi$ term in Eq.~\eqref{rho_phi}. In panels (b)--(c), each color identifies one $\xi$; solid lines precede the saved $\epsilon_V=1$ endpoint and dashed lines show the imposed post-inflationary dilution. Diamond markers in panels (d)--(f) denote the continuity-approximation calibration values.}
    \label{fig:EMOD}
\end{figure*}

In this subsection, we report the result of the numerical routine previously described.
The fitted parameters and adopted observable bounds are reported respectively in Tables~\ref{tab:fitted_parameters} and~\ref{tab:observable_bounds}. Notably, we calibrated the parameters for each value of $\xi$ and each potential choice. In this way, each model is addressed separately. 
The results in Figs. ~\ref{fig:STAR}--\ref{fig:TMOD} compare the GEF evolutions and the dependence of the CMB observables on $N_*$. 
Table~\ref{tab:magnetogenesis} gives the magnetic outputs under the stated post-inflationary assumptions. 

\begin{table*}[t]
\centering
\caption{Magnetogenesis outputs at the first rising $\epsilon_V=1$ event: inflationary duration $N_e$, total GEF magnetic density $\rho_{Be}$, and present-day root mean squared field $B_0$. The latter is obtained from the diffusion-filtered Fourier-mode spectrum assuming $65$ post-inflationary e-folds, a physical diffusion cutoff of $10^{-46}$ in reduced-Planck units, and no additional losses. The conversion is $B_0[{\rm G}]=3.04\times10^{56}\sqrt{2\rho_{B0}}$,  with $\rho_{B0}$ in reduced-Planck units. These conditional estimates are not obtained by redshifting the unfiltered $\rho_{Be}$ column. Negative-coupling rows do not satisfy the adopted CMB fit.}
\begin{tabular}{crccc}
\hline
$\qquad$Model $\qquad$& $\xi\qquad$ & $\qquad N_e\quad$ & $\quad\rho_{Be}\,[M_{P}^4]\quad$ & $\qquad B_0\,[{\rm G}]\qquad $\\ \hline 
Starobinsky & $-5\times10^{-4}$ & 60.000 & $1.45\times10^{-23}$ & $4.17\times10^{-22}$\\
Starobinsky & $0$ & 59.726 & $1.54\times10^{-23}$ & $7.17\times10^{-24}$\\
Starobinsky & $5\times10^{-4}$ & 59.938 & $1.14\times10^{-21}$ & $2.06\times10^{-23}$\\
$E$-model & $-5\times10^{-4}$ & 59.998 & $2.82\times10^{-23}$ & $2.49\times10^{-21}$\\
$E$-model & $0$ & 59.713 & $2.56\times10^{-23}$ & $1.21\times10^{-23}$\\
$E$-model & $5\times10^{-4}$ & 61.546 & $3.01\times10^{-22}$ & $2.10\times10^{-23}$\\
$T$-model & $-5\times10^{-4}$ & 64.476 & $1.41\times10^{-23}$ & $8.52\times10^{-21}$\\
$T$-model & $0$ & 59.712 & $2.38\times10^{-23}$ & $1.17\times10^{-23}$\\
$T$-model & $5\times10^{-4}$ & 74.706 & $9.10\times10^{-22}$ & $2.56\times10^{-23}$\\
\hline
\end{tabular}
\label{tab:magnetogenesis}
\end{table*}
\subsubsection{Starobinsky potential}
Fig.~\ref{fig:STAR} shows that the positive-coupling configuration leaves the inflationary plateau earlier in cosmic time and produces stronger electric amplification than the minimally coupled configuration. Negative coupling delays this transition and suppresses the electric density. Differently, the magnetic density presents early amplification for positive coupling and more gradual growth for negative coupling, followed by dilution.

The scalar amplitude rises with $N_*$ for minimal and negative coupling, whereas the positive-coupling curve decreases towards the accepted region. Instead, for the spectral index, the negative-coupling curve remains below the adopted bounds, while the minimal and positive configurations satisfy the scalar constraints. The tensor ratio remains allowed throughout the sampled interval. Thus, the model for negative coupling does not provide an observationally viable configuration.

\subsubsection{$E$-model potential}
The density histories in Fig.~\ref{fig:EMOD} show the same ordering of the transition at the end of inflation. Positive coupling gives earlier and stronger electric amplification, while negative coupling delays the electric growth. The magnetic curves follow a different ordering again, emphasizing that magnetic production cannot be directly inferred from the electric density.

The positive-coupling scalar amplitude satisfies observational bounds approximately for all the $N_*$ range studied, whereas the minimal and negative cases approach the allowed values for larger values of $N_*$. Its spectral index approaches the central accepted region, while the negative-coupling tilt is too red with respect to allowed values. The tensor ratio is larger for positive coupling but stays below the adopted bound. The minimal and positive configurations are therefore the viable candidates, while the enhanced present-day magnetic estimate for negative coupling again provides incompatible results with the adopted observational constraints.

\begin{figure*}[t]
    \centering
    \includegraphics[width=1\linewidth]{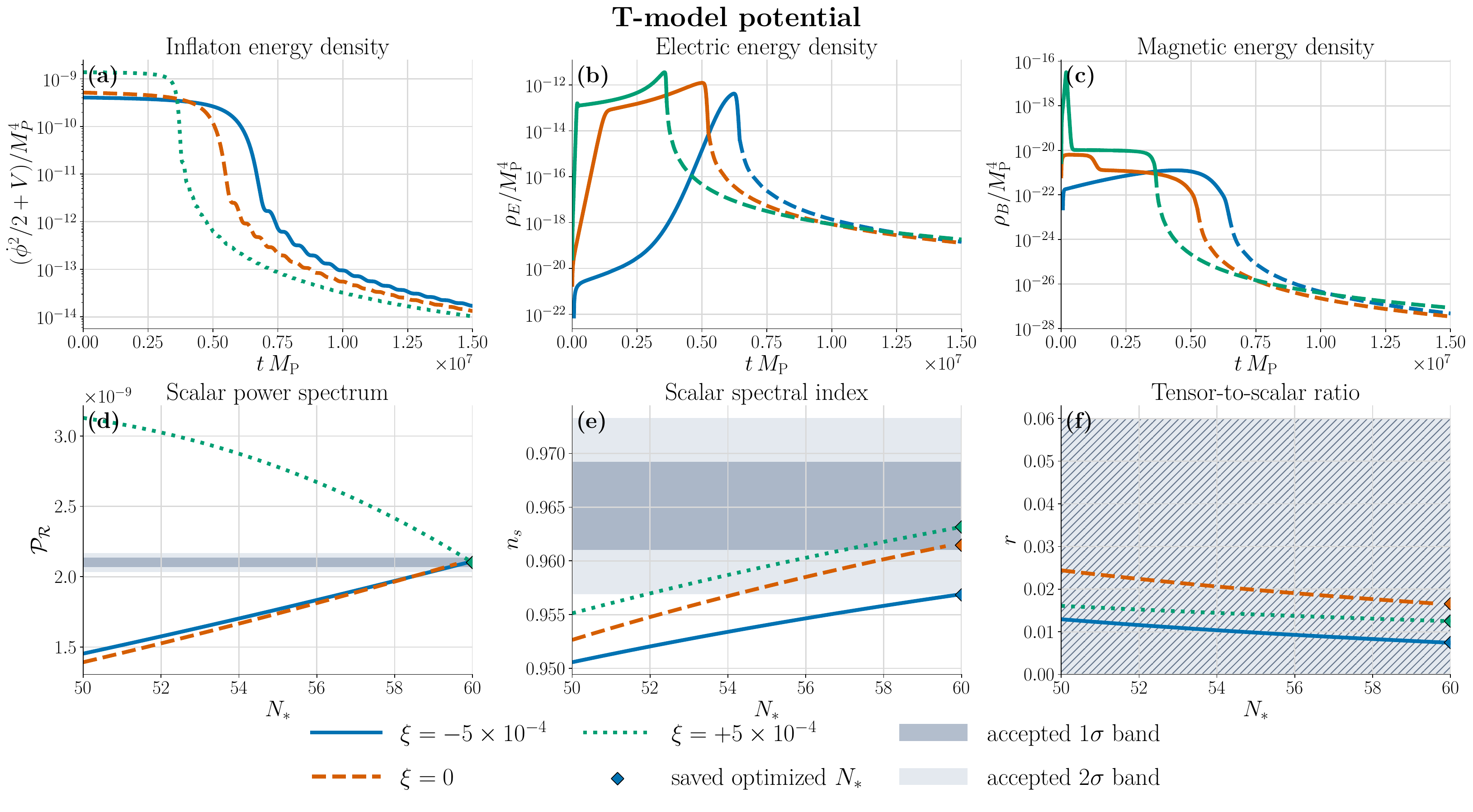}
    \caption{$T$-model potential with the potential-adapted coupling. Panels (a)--(c) show the inflaton, electric, and magnetic energy densities, respectively, as functions of time in reduced-Planck units. Panels (d)--(f) show the scalar curvature power spectrum $\mathcal P_{\mathcal R}$, scalar spectral index $n_s$, and tensor-to-scalar ratio $r$, respectively, as functions of the pivot-crossing e-fold number $N_*$. Panel (a) shows the canonical scalar contribution $\dot\phi^2/2+V$, excluding the $6\xi H\phi\dot\phi$ term in Eq.~\eqref{rho_phi}. In panels (b)--(c), each color identifies one $\xi$; solid lines precede the saved $\epsilon_V=1$ endpoint and dashed lines show the imposed post-inflationary dilution. Diamond markers in panels (d)--(f) denote the continuity-approximation calibration values.}
    \label{fig:TMOD}
\end{figure*}
\subsubsection{$T$-model potential}
Fig.~\ref{fig:TMOD} displays a pronounced electric growth also for negative coupling, followed by decay, while the late-time electric curves subsequently approach the same behavior. 

In this case, we observe that all the models corresponding to different $\xi$ values show a similar behavior: the $\PR$ observable approaches the allowed values for larger values of $N_*$ and exits the boundaries for smaller ones. The corresponding spectral index, instead, approaches the central accepted region for the positive and null $\xi$ case, while the negative-coupling tilt is too red with respect to allowed values. Negative coupling remains disfavored by the spectral index but yields the largest surviving magnetic estimate between the considered configurations. We remark that these exclusions regard the fitted points and sampled pivot interval, and not the potential families.

\section{Conclusions}\label{sec concl}

We investigated inflationary magnetogenesis with potential-dependent electromagnetic couplings in non-minimally coupled Starobinsky, $E$-model, and $T$-model inflation. Starting from a power-law scaling with the scale factor during slow roll, we derived explicit coupling functions adapted to the corresponding potentials. 
We then considered a parameter optimization with respect to CMB observables, with gradient expansion integrations including electromagnetic backreaction and Schwinger pair production. 
Finally, we computed the present-day magnitude of the produced magnetic field. 
In this regard, these estimates are qualitative since, in our analysis, we do not explicitly model the post-inflationary evolution, describing epochs such as reheating, and, mostly, because astrophysical phenomena may significantly affect the values obtained. 
In other words, the produced primordial magnetic fields have to be interpreted as seeds for the present-day abundance.

The fitted configurations show that the non-minimal coupling and the associated potential parameters affect both the timing of the inflationary transition and electromagnetic amplification. Positive-coupling configurations exhibit earlier electric amplification in cosmic time, while negative coupling generally delays its growth. These features do not directly determine the same behaviors for the surviving magnetic field, which depends on the spectral distribution and subsequent dilution. 
Among CMB observables, the scalar spectral index provides the strongest discrimination, indeed, the analyzed negative-coupling points remain too red, whereas minimally coupled and positive-coupling configurations can satisfy the adopted scalar and tensor bounds. Thus, the larger magnetic estimates associated with negative coupling do not provide viable magnetogenesis solutions. The minimal and positive configurations satisfy the adopted CMB constraints more consistently, but within this single prescription, their corrected magnetic amplitudes can not establish viable and complete explanations of intergalactic magnetic fields.
In any case, we remark that these statements hold for the considered configurations, and they do not exclude or validate an entire class of models within the entire range of the parameter space \cite{aggiunta1}.

The introduction of the new coupling function for the $E$- and $T$-model classes yields a well-defined generalization of the prescription originally obtained for the Starobinsky model. In particular, we showed that it does not apply only to standard general relativity, but also to non-minimally coupled magnetogenesis, thus extending the existing framework of inflationary magnetogenesis.

Future efforts will explore alternative and plausible versions of inflationary potentials, including exotic fields \cite{Belfiglio:2022qai,Luongo:2018lgy,DAgostino:2022fcx}, warm inflation \cite{DAgostino:2021vvv,Luongo:2025ovo,Luongo:2026ndg} and/or extended or modified theories of gravity \cite{Capozziello:2018hly,Belfiglio:2023rxb,Capozziello:2019cav,Abedi:2018lkr}. Additional studies will try to connect magnetogenesis to the expected deviations at recombination that allow to increase the  Hubble constant value, including early dark energy fields \cite{Poulin:2018cxd,Poulin:2021bjr} and alternatives to them \cite{Carloni:2025jlk,Carloni:2026yut}. Finally, we will clarify the role of spectator fields in the production of primordial magnetic fields \cite{Belfiglio:2024xqt}.

\section*{Acknowledgments}
The authors thank Orlando Luongo for valuable suggestions and feedback on inflationary magnetogenesis.

\bibliographystyle{unsrt} 
\bibliography{test}

\end{document}